\documentclass[runningheads]{llncs}
\usepackage[T1]{fontenc}
\usepackage{xcolor}
\usepackage{subfig}
\usepackage{algorithm}
\usepackage{algpseudocode}
\usepackage{amsmath}
\usepackage{amsfonts}
\usepackage{graphicx}
\begin{document}
\title{Leveraging band diversity for feature selection in EO data}
%
%
\author{Sadia Hussain\inst{1}\orcidID{0000-0002-5507-1063} \and
Brejesh Lall\inst{2}\orcidID{0000-0003-2677-3071}}
\authorrunning{S. Hussain et al.}
%
\institute{Bharti School of Telecommunication Technology and Management , IIT Delhi, Hauz Khas, New Delhi, India 
\email{bsz198107@dbst.iitd.ac.in}\\ \and
Electrical engineering department, IIT Delhi, Hauz Khas, New Delhi, India\\
\email{brejesh@ee.iitd.ac.in}}
\maketitle              
\begin{abstract}
Hyperspectral imaging (HSI) is a powerful earth observation technology that captures and processes information across a wide spectrum of wavelengths. Hyperspectral imaging provides comprehensive and detailed spectral data that is invaluable for a wide range of reconstruction problems. However due to complexity in analysis it often becomes difficult to handle this data. To address the challenge of handling large number of bands in reconstructing high quality HSI, we propose to form groups of bands. In this position paper we propose a method of selecting diverse bands using determinantal point processes in correlated bands. To address the issue of overlapping bands that may arise from grouping, we use spectral angle mapper analysis. This analysis can be fed to any Machine learning model to enable detailed analysis and monitoring with high precision and accuracy. 

\end{abstract}

\section{Introduction}
A hyperspectral imaging device collects spectral information using hundreds of narrow bands and combines it with digital imagery. When an object interacts with light at different wavelengths, this technology captures the unique physical and chemical characteristics of the material. As a result, hyperspectral imaging finds numerous applications in earth observation, including precision agriculture, climate monitoring, and remote sensing. However, this technology comes with several challenges. With information spread across a wide range of narrow bands, similar yet distinct objects can be mistakenly categorized as the same. Additionally, there is a bottleneck in transmitting these images when captured by a sensor, which often necessitates processing these extensive bands. This processing may degrade either the spatial resolution, the spectral resolution, or both. The large size of hyperspectral imaging (HSI) data introduces multiple processing issues, such as increased computational cost, complexity in image analysis, and a scarcity of available training data. Collecting large datasets can be difficult due to limitations of acquisition devices, and storing this vast amount of information can also be cumbersome. Consequently, in restoration-based problems, machine learning or computer vision-based solutions often prove inadequate at these early stages due to the need to process a large number of bands. In hyperspectral images, the high dimensionality results from the large number of bands. Effective hyperspectral band management is crucial for revealing the unique features of objects. Consequently, two main approaches are commonly found in the literature: band selection (or feature selection) and band extraction (or feature extraction). Band selection involves using significant or representative bands. Based on the physical and chemical characteristics of the material, these representative bands are selected to compactly represent hyperspectral images. On the other hand, band extraction reduces the higher dimensionality of hyperspectral bands into a lower dimensionality, which can cause hyperspectral images to lose their physical significance. 
In this position paper, we devise a unified band selection approach for controlling hyperspectral bands by applying a grouping strategy. This grouping strategy enhances the performance of any machine learning-based method used for resolution enhancement. Our paper outlines three important areas of grouping diversity by employing: (1) Sampling Method: We use Determinantal Point Processes to provide insights for the selection of diverse bands. (2) Spectral Correlations: We group strongly correlated bands together based on their spectral characteristics. (3) Spectral Angle Mapping analysis: For overlapping bands within a group, we use the spectral angle mapper to disentangle bands based on more precise similarity measurements.

This unified approach aims to optimize the use of hyperspectral bands, improving the accuracy and efficiency of hyperspectral image analysis in various applications.
\begin{figure}
\includegraphics[width=\textwidth]{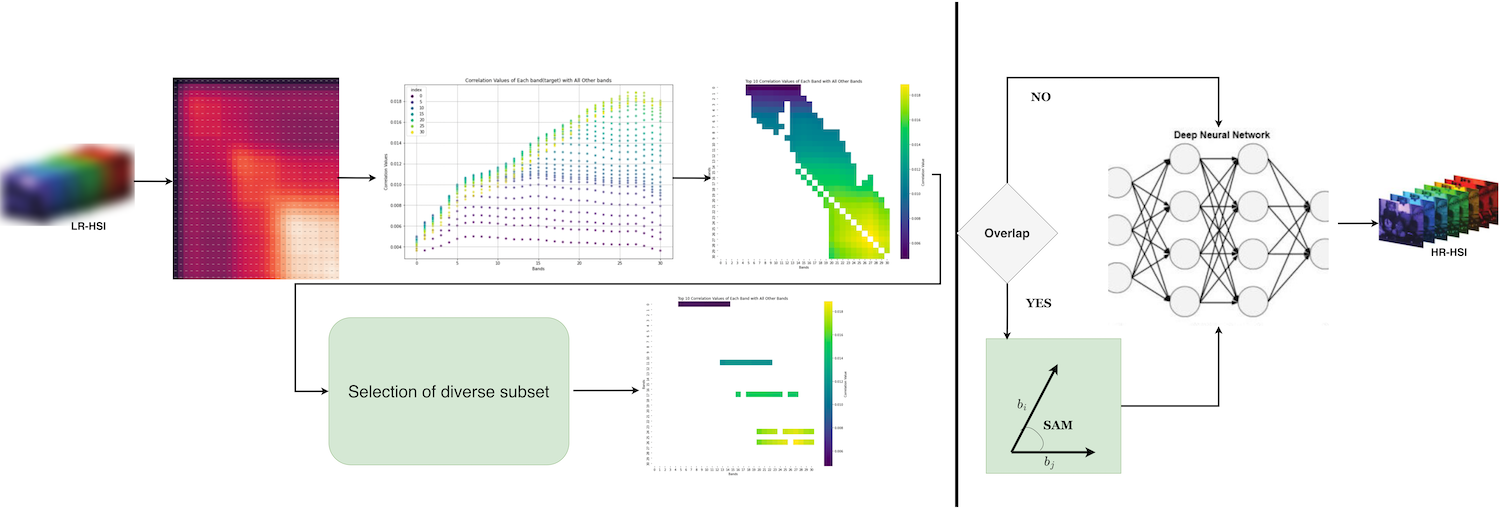}
  \caption{Schematic view of the proposed approach. }
  \label{fig:FIG2}
\end{figure}
\section{Related Works}
A lot of literature is found in Image classification using the above two approaches. In band selection methods further three sub-categories have been devised based on the derivation of subset of bands~\cite{sawant2020survey}. These are as follows: subsets derived on the basis of subset evaluation criteria~\cite{chang1999joint,tschannerl2019mimr,bhardwaj2018unsupervised,keshava2004distance,zhang2018fast}, availability of prior information further sub-categorized on the basis of supervised selection criteria~\cite{guo2014improving,yang2010efficient}, unsupervised selection criteria~\cite{sawant2020unsupervised,jia2012unsupervised} and selection strategy (individual~\cite{datta2015combination,jia2015novel} and other evaluation techniques) used to create the band subset.

Efficient band selection plays a crucial role in extracting meaningful information from vast datasets. By selecting a subset of relevant spectral bands, researchers can reduce data dimensionality, enhance computational efficiency, and improve the performance of downstream analysis tasks such as restoration tasks, classification and target detection. However, achieving optimal band selection poses significant challenges, necessitating innovative approaches that leverage spectral grouping techniques.~\cite{9153939},~\cite{10036355} address this need with novel methodologies. The former introduces a method based on neighborhood grouping to efficiently identify relevant bands, while the latter proposes leveraging differences between inter-groups for band selection.~\cite{ayna2023learning} presents a learning-based optimization approach for band selection tailored specifically for classification tasks. Despite their innovative approaches, both approaches face challenges interms of computational efficiency, interpretability, and scalability. Furthermore, in the context of hyperspectral image restoration, particularly superresolution, no definitive solutions have emerged regarding grouping methodologies. This underscores a critical gap in current research highlighting the need for further exploration and development in this area. Addressing these challenges is crucial for an efficient grouping algorithm, advancing hyperspectral imaging capabilities and realizing the full potential of band selection techniques in various applications.
\begin{figure}[h]
    \centering
    \subfloat{\includegraphics[width=0.20\textwidth]{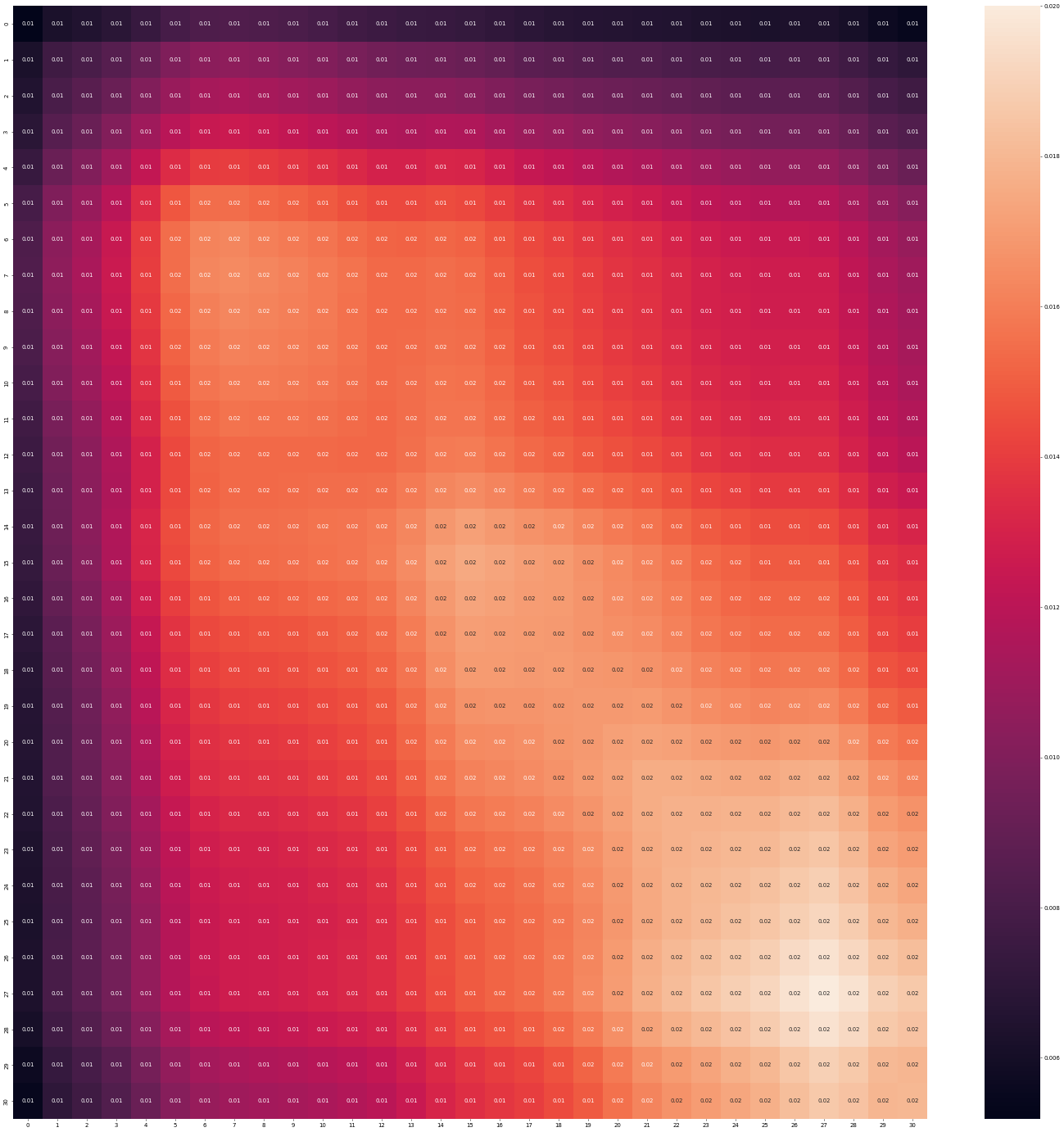}} 
    \subfloat{\includegraphics[width=0.20\textwidth]{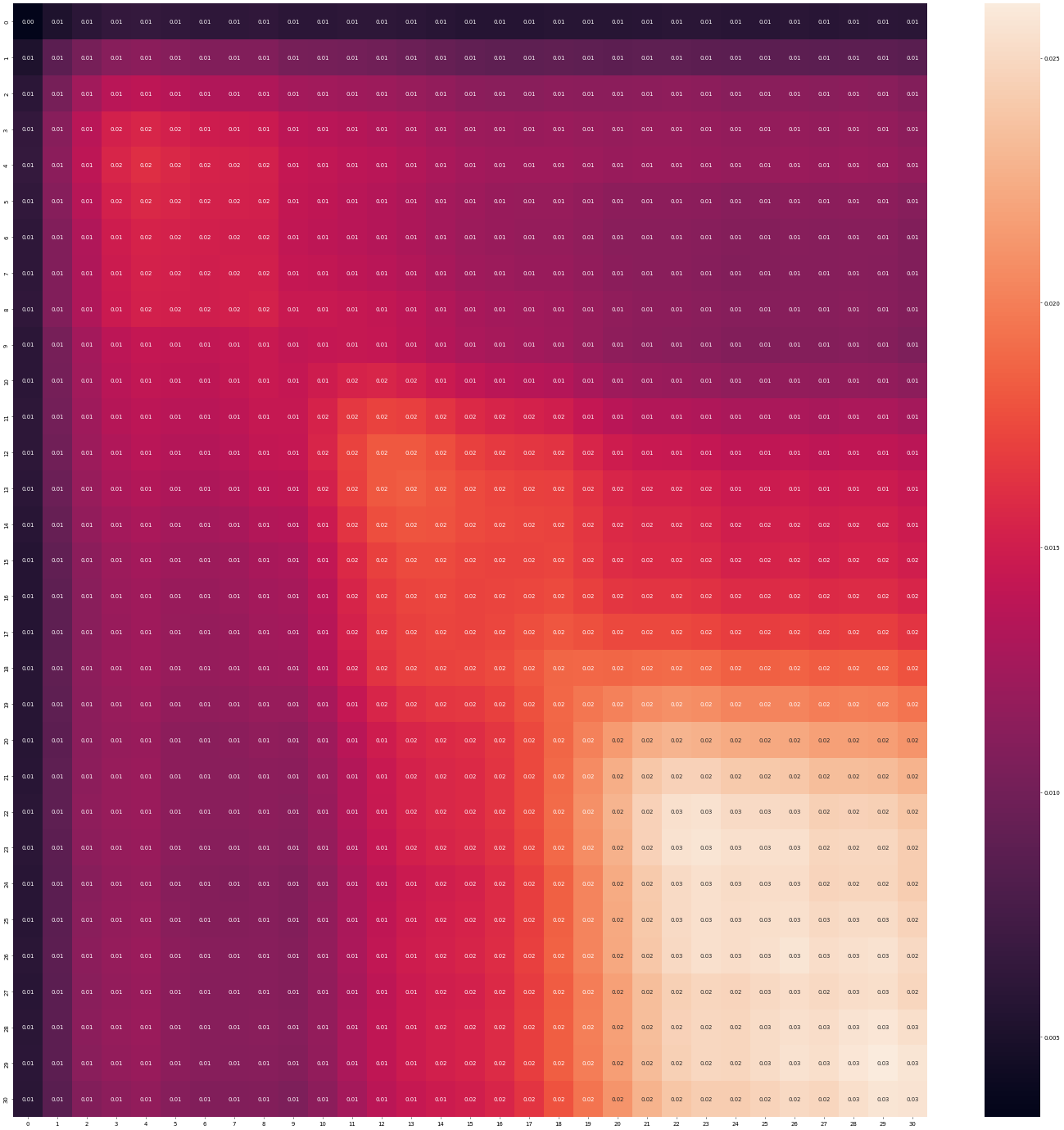}} 
    \subfloat{\includegraphics[width=0.20\textwidth]{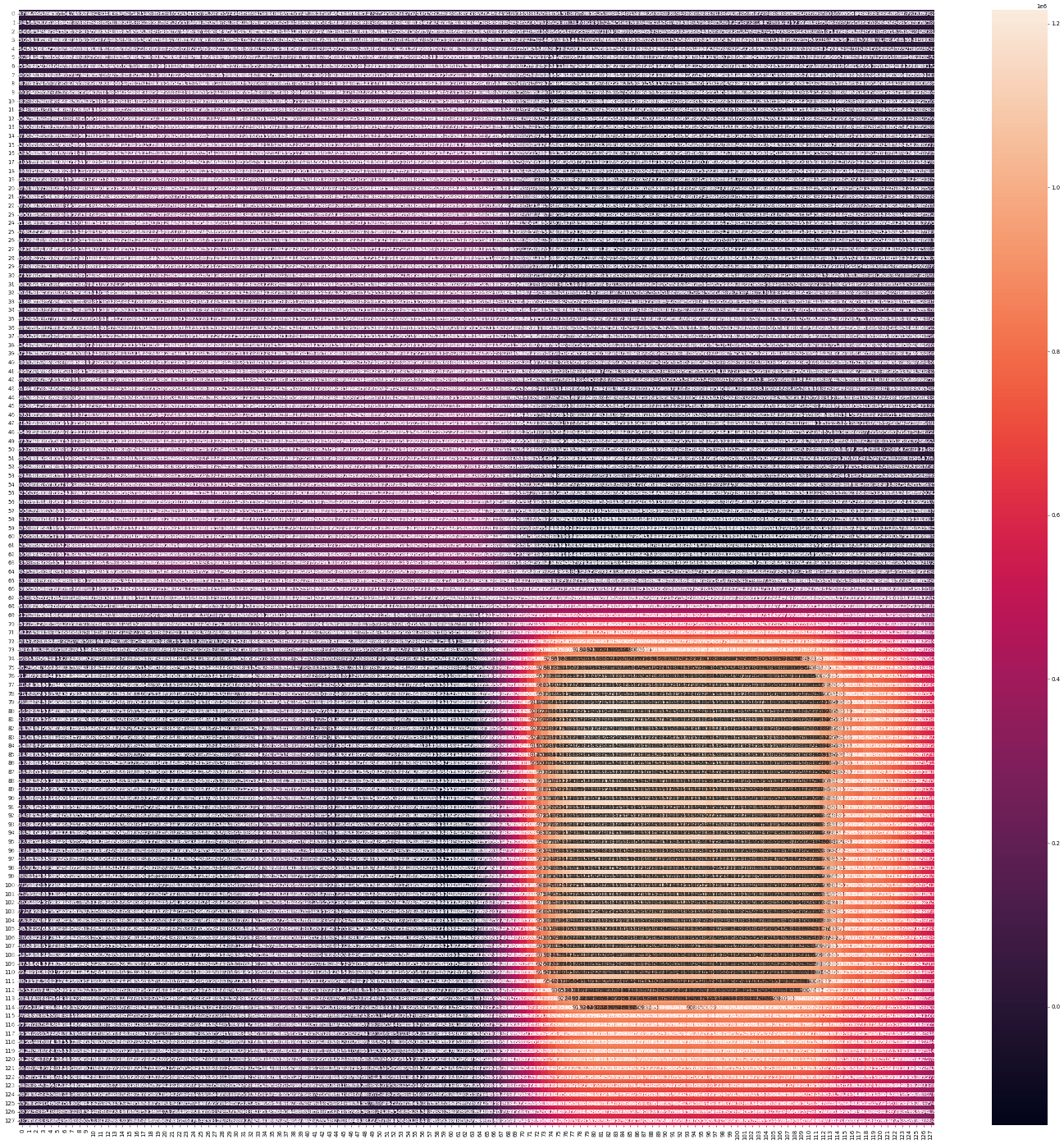}}
    \subfloat{\includegraphics[width=0.20\textwidth]{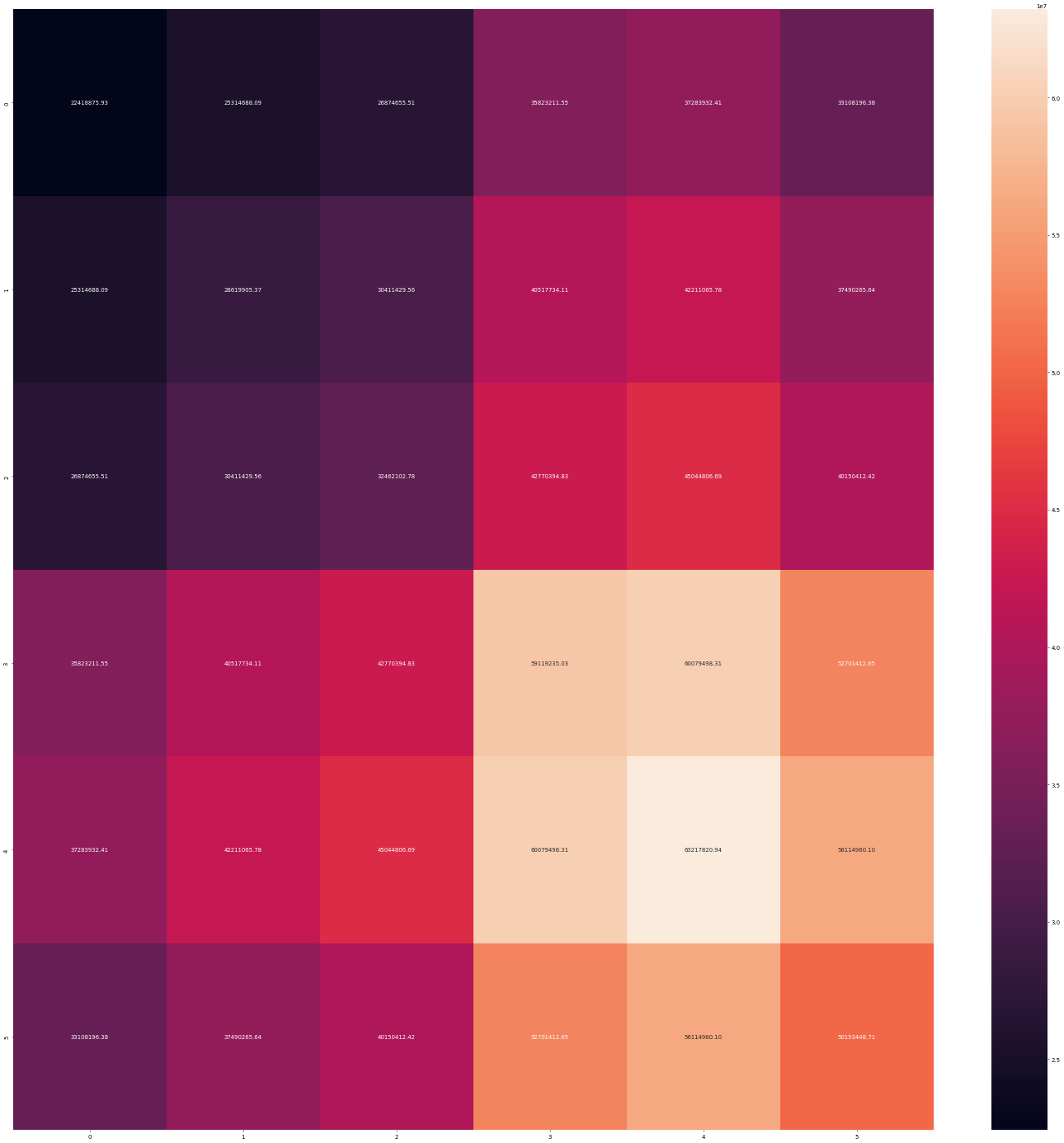}}
    \subfloat{\includegraphics[width=0.20\textwidth]{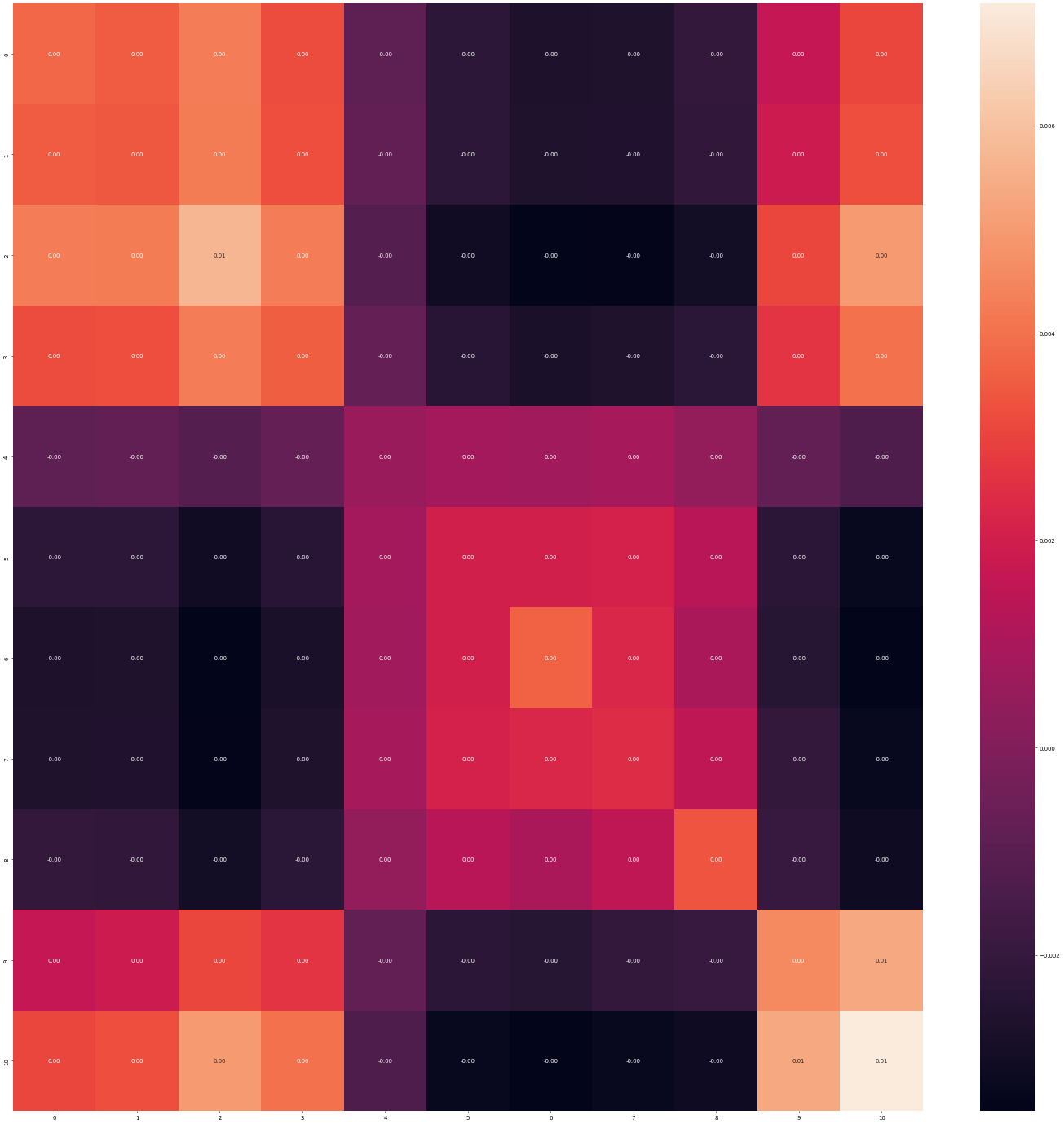}}
    \caption{(a) Correlation matrix for NTIRE2022 Dataset (b) CAVE Dataset (c) Chikusei Dataset (d) Sentinel Dataset (e) Landsat Dataset}
    \label{fig:foobar}
\end{figure}
Recent advancements in hyperspectral image analysis focus on correlation matrices, which reveal spectral dependencies and enable high-texture detail within groups of spectrally dependent vectors. Traditionally, linear predictions~\cite{manolakis2008spectral} have dominated hyperspectral data analysis, but recent studies advocate for interval sampling~\cite{wang2022group} for superior group formation and enhanced interpolation outcomes. Our study proposes an explicit grouping method based on correlation coefficients to establish a standardized framework for hyperspectral super-resolution. By integrating correlation analysis into our grouping strategy, each band within a correlated group is rigorously evaluated for its significance in super-resolution. These explicit groups, guided by correlation analysis, feed into deep neural network architectures, optimizing interpolation learning efficiency. Our innovation lies in combining Determinantal Point Process (DPP) with correlation analysis to form coherent band subsets, minimizing redundancy and maximizing diversity. This integration enriches interpolation learning by capturing intricate spectral relationships effectively. Challenges arise when grouped bands still overlap significantly despite spectral correlation. To mitigate this, our approach employs Spectral Angle Mapper (SAM) to resolve overlaps based on the lowest SAM values, refining the grouping strategy. This modular integration enhances cohesion, learning robustness, and efficiency in handling complex hyperspectral data, promoting seamless interaction within the network and improving restoration outcomes.
\begin{algorithm}
\caption{Spectral correlation estimation}
\label{algo: Algorithm1}
\begin{algorithmic}[1]
\State \textbf{Input:} \(X\), \(S\)
\State \textbf{Output:} \(Z\)
\State \(Z \in \mathbb{R}^{N \times WH}\)
\State \(X \in \mathbb{R}^{N \times wh}\)
\State Initialize \(B \in \mathbb{R}^{N \times n}\)
\State Initialize \(M \in \mathbb{R}^{n \times WH}\)
\State \(X = ZS\)
\For{\textbf{each band} \(i, j\)}
    \State \(b_i = f_i(t)\)
    \State \(b_j = f_j(t)\)
    \State \(R_{b,b}(i, j) = \frac{E[b_i, b_j]}{\sigma_i \sigma_j}\)
\EndFor
\State \(Z = B \cdot M\)
\State Solve for \(B\) and \(M\)
\end{algorithmic}
\end{algorithm}
\section{Methodology}
\begin{algorithm}
\caption{k-DPP Sampling}
\label{algo: Algorithm 2}
\begin{algorithmic}[1]
\State \textbf{Input:} Hyperspectral data $L$
\State \textbf{Output:} Subset of bands $Y$ of size $k$
\State Initialize \( k \)-DPP for sampling band subsets
\State Set cardinality constraint $k$
\State Compute the probability $P^k_L(Y)$ of selecting a subset $Y$ of size $k$ from $L$ using k-DPP:
\[
    P_{L} (\mathbf{Y} = Y ) = \frac{\det(L_{Y})}{\sum_{\mathbf{Y} \subseteq \gamma} \det(L_Y)} = \frac{\det(L_{Y})}{\det(L + I_{N})}
\]
\State Decompose $L_{Y}$ into eigenvectors $S$ and eigenvalues $\lambda_{n}$
\State Select subset of eigenvectors based on eigenvalues:
\[
    P(n \in S) = \frac{\lambda_{n}}{e_{k}^{n}} \sum_{|S'| = k-1} \prod_{n' \in S'} \lambda_{n'} = \lambda_{n} \frac{e_{k-1}^{n-1}}{e_{k}^{n}}
\]
\State Ensure selected subset represents the most relevant and informative components
\State \textbf{return} Subset of bands $Y$ of size $k$
\end{algorithmic}
\end{algorithm}
This section elaborates our proposed approach for Band grouping. Our method comprises of three primary components for grouping bands in a data: extracting primary grouping information based on correlation analysis, extracting critical band information using DPP, solving for overlapping bands using spectral angle mapper information. The technical specifics of these components are elucidated in the following discussion.
\subsection{Spectral Correlation in HSI}
Correlation functions in hyperspectral imaging provide valuable insights into the relationships between spectral bands and spatial locations in hyperspectral data. These correlation functions can help understand how the spectral characteristics of different bands are related and can be used for various purposes, including feature selection, dimensionality reduction, and image interpretation. 
As illustrated in Algorithm~\ref{algo: Algorithm1}, we denote the high-resolution hyperspectral image (HR-HSI) data as \( Z \in \mathbb{R}^{W \times H \times N} \) and the low-resolution counterpart (LR-HSI) as \( X \in \mathbb{R}^{w \times h \times N} \). The LR-HSI \( X \) is obtained from \( Z \) through spatial downsampling using matrix \( S \) such that \( X = ZS \), where \( S \in \mathbb{R}^{WH \times wh} \). To recover \( Z \), which captures the spatial and spectral details of the original HR-HSI, we propose a method involving spectral basis \( B \in \mathbb{R}^{N \times n} \) and a correlation matrix \( M \in \mathbb{R}^{n \times WH} \): \( Z = BM \). Here, \( B \) represents spectral basis vectors, and \( M \) incorporates correlation coefficients. This formulation aims to find optimal values for \( B \) and \( M \) to reconstruct \( Z \). Understanding the spatial and spectral correlations inherent in hyperspectral images is crucial. Bands that are closer together often exhibit similar patterns due to underlying scene properties. This correlation structure is assessed through auto-correlation measures \( R_{b,b}(i, j) \) in Algorithm 1.

As illustrated in Algorithm 1, each element in \( R_{b,b}(i, j) \) represents the correlation between band \( i \) and band \( j \). This analysis aids in identifying which bands carry similar information essential for feature selection. A higher value indicates that these bands change in a similar manner across pixels, suggesting they may capture similar spectral information. Conversely, a lower value or near-zero covariance indicates that the bands change independently.
\subsection{Determinantal Point Processes}
Our proposed method extends the original k-DPP approach~\cite{kulesza2011k,kulesza2012determinantal} for sampling diverse band subsets from hyperspectral data, represented by the correlation matrix $L$ as illustrated in Algorithm~\ref{algo: Algorithm 2}. It ensures a subset of size $k$ with diverse bands via the probability $P^{k}_{L}(Y)$, which conditions on subsets' diversity captured by $L_{Y}$ and the eigenvalues $\lambda_{n}$, ensuring relevance and diversity in the selected subsets. Eigenvectors $S$ are selected based on eigenvalues $\lambda_{n}$ to capture dataset variability effectively.
\subsection{SAM for overlapping Bands}
In the context of grouping strategies in HSI, SAM is employed to handle overlapping bands within a group. Overlapping bands can complicate the analysis as they may contain redundant information or obscure the distinct spectral features. By applying SAM as illustrated in Algorithm 3, we can measure the precise similarity between bands within a group and disentangle those that are too similar.
\begin{algorithm}
\caption{Calculate Spectral Angle Mapper (SAM)}
\begin{algorithmic}[1]
\Function{CalculateSAM}{$\vec{v}_1, \vec{v}_2$}
    \State $numerator \gets \vec{v}_1 \cdot \vec{v}_2$ \Comment{Dot product of vectors}
    \State $denominator \gets \|\vec{v}_1\| \cdot \|\vec{v}_2\|$ \Comment{Product of norms}
    \State $angle \gets \arccos\left(\frac{numerator}{denominator}\right)$ \Comment{Calculate angle}
    \State \Return $angle$
\EndFunction
\State $diverse\_set \gets \{ k\}$ \Comment{Diverse set elements}
\State $sam\_values \gets \mathbf{0}_{|diverse\_set| \times 31}$ \Comment{Initialize SAM values matrix}

\For{$i \in \{1, \dots, |diverse\_set|\}$}
    \For{$j \in \{1, \dots, 31\}$}
        \If{$diverse\_set[i] \neq j$}
            \State $sam\_values[i, j] \gets \text{CalculateSAM}(band\_data_1, band\_data_2)$
        \Else
            \State $sam\_values[i, j] \gets \text{NaN}$
        \EndIf
    \EndFor
\EndFor
\end{algorithmic}
\end{algorithm}
\section{Conclusion}
In conclusion, our proposed approach, which utilizes Determinantal Point Processes (DPP) and spectral angle mapping (SAM), offers a promising direction for addressing the complexities of hyperspectral imaging (HSI). By applying these methodologies, we aim to enhance the effectiveness of spectral band selection and optimize hyperspectral image reconstruction. This approach seeks to mitigate redundancy while striving to improve the efficiency and accuracy of analysis techniques. Future research into advanced grouping strategies will be essential for overcoming remaining computational and interpretative hurdles, thereby unlocking the full potential of HSI in various earth observation applications.
\bibliography{samplepaper}
\bibliographystyle{splncs04}
\end{document}